# Adhesion forces due to nano-triboelectrification between similar materials

C. Guerret-Piécourt*[1,3], S. Bec[2], F. Ségault[2], D. Juvé[3], D. Tréheux[3], A. Tonck[2]

[1] Laboratoire de Physico-Chimie des Polymères, CNRS UMR 5067, LPMI-CURS, BP 1155-64013 PAU Cedex, France – Since January 1st 2006 at Laboratoire de Tribologie et Dynamique des Systèmes, UMR 5513 CNRS / ECL / ENISE, Ecole Centrale de Lyon, BP 163, 69134 ECULLY Cedex
[2] Laboratoire de Tribologie et Dynamique des Systèmes, UMR 5513 CNRS / ECL / ENISE, Ecole Centrale de Lyon, BP 163, 69134 ECULLY Cedex
[3] Laboratoire d'Ingénierie et Fonctionnalisation des Surfaces, CNRS UMR 5621, Ecole Centrale de Lyon, BP 163, 69134 ECULLY Cedex

*Corresponding author:
Christelle GUERRET
Tel: 33-5-59407708, Fax: 33-5-59407744, e-mail: christelle.guerret@univ-pau.fr
<u>Since January 1st 2006:</u>
Tel: 33-4-72186301, Fax : 33-4-78433383, e-mail : christelle.guerret@ec-lyon.fr

**Abstract**. Contact electrification and triboelectrification are well-known in the case of dissimilar materials, however the case of charge exchange during friction between nominally identical insulating materials is less documented. We experimentally investigated the triboelectrification between two smooth monocrystalline α-$Al_2O_3$ (sapphire) antagonists by surface force measurements with a Surface Force Apparatus (SFA). The force between a sphere and a plane, both in sapphire, was measured as a function of the sphere-plane distance D, before and after nano-friction tests, under dry argon atmosphere. Respective contributions of van der Waals, water meniscus and electrostatic forces were determined. The estimated Hamaker constant was in good agreement with the Lifshitz theory, and the dominant meniscus attraction at low separation could be overcome with small radius sphere. We demonstrated that electrostatic forces were generated by the nano-friction test and we quantified the adhesion that results from this contact-electrification. In the first stage of the unloading process, the short range electrostatic force was found to vary both with time and distance D. Experimental results were correlated with surface densities of mobile charges on the two surfaces, and the time-dependence was related to classical surface transport phenomena on alumina surfaces.

## 1. Introduction

The phenomenon of electrical charging, when two materials are brought into contact or rubbed together, is a well-known effect, whose consequences on adhesion are very important [1, 2]. Generally, triboelectrification is evidenced between two dissimilar materials. In the case of metal-metal contact, it has been attributed to an electron transfer until the two Fermi levels equilibrate [3]. For metal-insulator, or insulator-insulator contact or friction, no satisfactory physical explanation of this phenomenon has been established despite the great number of published studies on this topic, as concluded by Lowell and Rose-Innes in their review paper [4].

Thanks to the recent progress in atomic force microscopy (AFM), the elucidation of contact electrification or triboelectrification, in the case of metal-insulator or insulator-insulator contacts, knows a renewed interest [5, 2, 6-8]. These studies are mainly dedicated to the metal-insulator case, because AFM tips are usually tungsten or platinum tips, and it is demonstrated that the long range interaction due to the electrostatic force contribute to an important part of the work of adhesion.

The case of triboelectrification, between two nominally identical materials, has been less studied despite its theoretical and applicative interests. Some experiments have been made at the macroscopic scale, and different theories have been proposed for the charge transfer associated to asymmetry in the rubbing [9]. For instance, physical difference between the two surfaces (in particular temperature difference) or transfer resulting from non-equilibrium electron distributions have been proposed as tentative mechanisms.

To our knowledge, no evidence of this electrification phenomenon of similar materials has yet been demonstrated at the nanometric scale, except in the case of one of the two surfaces coated with a single chemisorbed monolayer [10]. The aim of this paper is to show evidence of the generation of electrical charges during the nanofriction of two sapphire surfaces under dry argon atmosphere.

## 2. Experimental considerations

For this study, we used a sphere-plane contact. The radius of the macroscopic sphere is about 3 mm. The surface forces between the two solids, both in sapphire, were measured versus the separation distance using a surface force apparatus.

### 2.1 Experimental set-up

*2.1.1 Material*

Monocrystalline $\alpha$-alumina is a good candidate for such a study at nanometric scale for several reasons. (i) Because sapphire surfaces can be easily polished, it avoids usual experimental difficulties linked to surface roughness. So, the two solids (sphere and plane) were carefully polished so that their mean roughness did not exceed a few nanometers and allowed accurate surface force measurements. (ii) The second reason is related to the amount of traps for electrical charges inside sapphire. Indeed, most of the theories about the electrical charging of insulators involve the electronic states available at their surfaces [7, 9]. In the case of sapphire, it has been demonstrated in previous studies using the Scanning Electron Microscope Mirror Effect method (SEMME) [11], that the density of defects was directly related to the density of available states for electrons. Considering these parameters, the sapphire sphere and plane were annealed at 1000°C during 4 hours in air in order to control the quantity of structural defects (particularly the amount of dislocations induced by polishing and of oxygen vacancies). This annealing temperature has been demonstrated to lead to the maximum ability of charge trapping, i.e. to the maximum of available states, at room temperature [12].

Prior to their mounting in the SFA, samples were cleaned with a sulfo-chromic solution to minimize surface contamination [13]. XPS analysis on the cleaned surface has shown that no conducting element (residual chromium) remained on the solid surfaces. Before introducing argon in the SFA chamber, vacuum was maintained at $6 \cdot 10^{-5}$ mbars during 24 hours to minimize the occurrence of water.

Two types of spheres were used with and without asperities at their surface. Experiments of type (I) were performed with perfectly polished (mean roughness less than few nanometers) sphere and plane. In experiment of type (II), the mean roughness is still weak but there was an isolated asperity on the sphere surface that was located in the contact area.

*2.1.2 Experiments*

The Ecole Centrale de Lyon Surface Force Apparatus (SFA) used in this study has been precisely described in previous publications [14, 15]. A schematic diagram of the device is shown on figure 1. The general principle is that a macroscopic spherical body -or a diamond tip- can be moved toward and away from a planar one using the expansion and the vibration of a piezoelectric crystal along the three directions, Ox, Oy (parallel to the plane surface) and Oz (normal to the plane surface). The plane specimen is supported by double cantilever sensors, measuring normal and tangential forces. The sensor's high resolution (10 nN) allows a very low compliance to be used for the force measurement (up to $2 \ 10^{-6}$ m/N). Three capacitive sensors were designed to measure relative displacements in the three directions between the supports of the two solids, with a resolution of 0.01 nm in each direction. Each sensor capacitance is determined by incorporating it in an LC oscillator operating in the range 5-12 MHz. Experiments are conducted either by imposing a displacement to the sphere and measuring the resulting forces, or by imposing a force to the sphere and measuring the resulting displacement. Dynamic measurements are simultaneously obtained by adding small vibrations to the quasi-static slow movement both in normal and in tangential directions. So normal and tangential stiffness are obtained from the in-phase signals, and the out of phase signals are related to the dissipative phenomena, such as viscous damping.

This instrument permits several kinds of experiments performed in controlled atmosphere: sphere/plane experiments (nano-rheology experiments, nano-friction tests) or tip-plane experiments (nano-indentation) [16]. In this study, the SFA was used to perform surface force measurements before and after nano-friction experiments, in dry argon atmosphere.

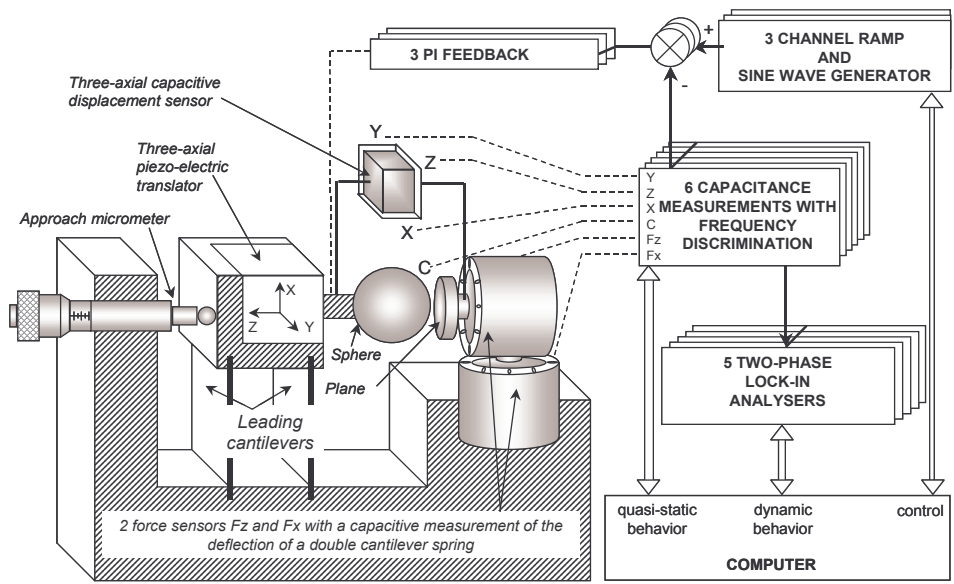

*Figure 1:* *Schematic diagram of the three-axis Surface Force Apparatus (SFA).*

## 2.2 Description of a typical test with the SFA

Figure 2 shows the imposed normal displacement Z and the resulting normal load $F_z$ versus time for one representative experiment of type (II) (asperity on the sphere surface located in the contact area).

This test is rather complex because the displacement includes different steps:

(ABC) and (CDE): 2 successive loading-unloading cycles, at slow imposed normal speed (a few Angstroms per second), with a maximum load of 100 µN and without nano-friction,
(EF): loading test at slow imposed normal speed with a maximum load of 1000 µN,
(FG): nano-friction alternative test of 128 cycles, 600 nm long, at imposed load of 1000 µN,
(GH): unloading test at slow imposed normal speed. During unloading, the displacement is stopped (S) during a few minutes to record the evolution of the load versus time,
(HIJ) and (JKL): successive loading-unloading cycles, similar to (ABC) and (CDE).

During the tests, continuous dynamic measurements are conducted using the vibration of the piezoelectric crystal. The normal and tangential stiffnesses (related to the elasticity of the solids) and the damping function (related to dissipative processes such as viscous behaviour) are recorded simultaneously with the quasi-static load and displacement measurements. The zero for the displacement is placed where both the normal load and the normal and tangential stiffnesses begin to increase (solid contact between the sphere and the plane).

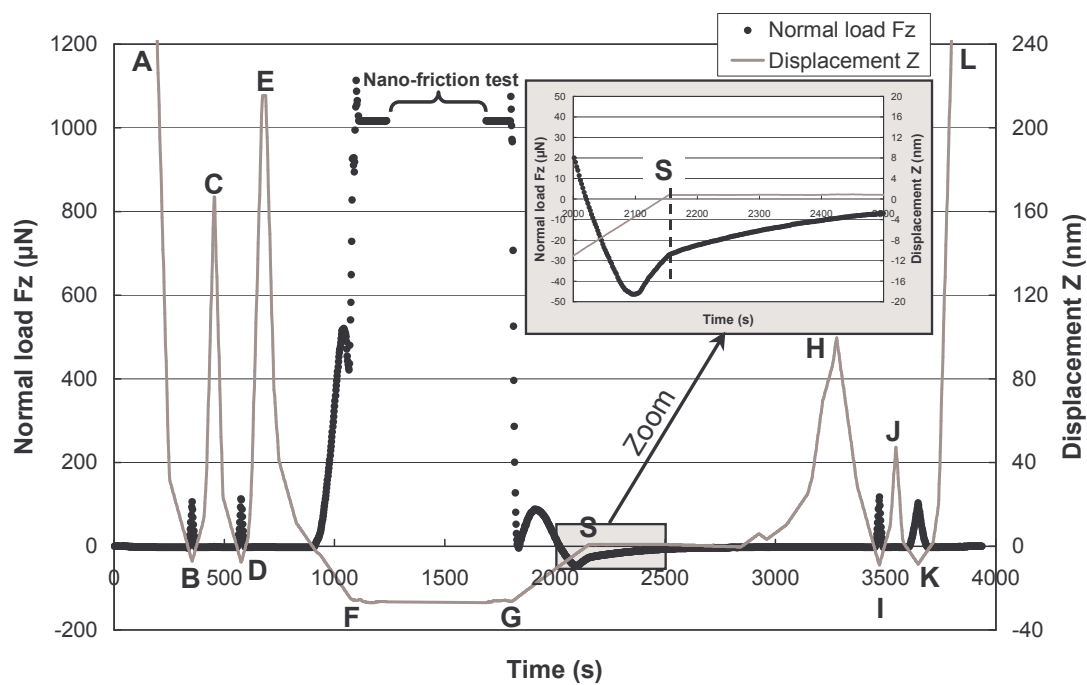

*Figure 2*: Imposed normal displacement Z for a representative type II test (grey curve) and resulting normal load $F_z$ versus time curve (black dots). Inset corresponds to a zoom of the normal load $F_z$ versus time curve.

## 3. Results and discussion

### 3.1 Surface force measurement results

*3.1.1 Generalities about surface forces between two solids.*

The surface forces involved between two solids in contact result from the contribution of the interatomic forces between all the atoms of the system. They include those of the two bodies plus those of the medium surrounding them. These surface forces are described in details by Israelachvili in [1] and can be classified according to their ranges:

(i) Very-short-range forces result from the exchange of electrons like in covalent bonding or hydrogen bonding.

(ii) The van der Waals interactions are short-range forces due to various interactions between atomic or molecular electric dipoles. Different approaches have been developed to calculate them. The Hamaker's model is based on a simple pairwise additive model, which ignores the influence of neighbouring atoms on the interaction between any pair of atoms. Combine with the Derjaguin approximation, it is frequently used to calculate simply the van der Waals force between a sphere and a plane, which is then proportional to reverse of the square of the sphere-plane distance. The Lifshitz theory ignores the atomic structure, and the forces between bodies, considered as continuous media, are expressed from bulk properties like dielectric constants and refractive indexes.

In the presence of water vapor in the medium surrounding the two solids, additional forces have to be taken into account like the capillary force due to the formation of a water meniscus. The range of this force is limited by the break of the meniscus, and the measurement of viscous damping is characteristic of the presence of a meniscus.

(iii) The coulombic interaction is the stronger and long-range surface force. In the case of solids immersed in a polar environment the formation of an electrical double layer can be described by the well-known DLVO theory. However, in non-polar environments (vacuum, dry nitrogen...) long-range coulombic forces during friction between two insulating materials become significant.

*3.1.2 Preliminary results from experiments of type (I)*

Experiments of type (I), with very smooth plane and sphere surfaces, were used to verify the features of the contact between the two sapphire surfaces. A small attraction between the two surfaces could be measured for small separation distance D. Indeed this attraction was adequately fitted typically between 5 and 15 nm by a non-retarded van der Waals interaction (slope equal to minus two, for log-log plot $F_z(D)$ where Fz is the normal force). An example of measured van der Waals attraction is shown on figure 3. The corresponding Hamaker constant H can be deduced from the equation $F_z/R=-H/6D^2$, deriving from the Derjaguin approximation in the case of sphere-plane contact (R is the radius of the sphere). We find $7 \cdot 10^{-20}$ J, which is between the literature values (obtained from the Lifshitz theory) for sapphire surfaces in air $H_{sapphire/air/sapphire}= 14 \cdot 10^{-20}$ J and in water $H_{sapphire/water/sapphire} = 5.3 \cdot 10^{-20}$ J [1]. This suggests that we did not succeed in removing water vapor despite the special experimental cares taken to eliminate it (vacuum of $6 \cdot 10^{-5}$ mbars, and dry argon atmosphere during the experiment). The presence of water on the surfaces was also confirmed by the appearance of viscous damping at smaller separation distance (revealed by the increase of the out of phase signal ωAz, see example on figure 2), which is characteristic of the formation of a meniscus.

*3.1.3 Surface force measurements from experiments of type (II).*

We first focus on the step (EF) of the test (as defined on figure 2), that corresponds to the approach of the two surfaces at slow imposed normal speed with a maximum load set to 1000μN. Figure 4 shows typical dynamic signals during this step. From both the load and the stiffness measurements, we observed that, for the 100 μN cycles and until the load reaches 500 μN, the contact does not correspond to the contact of a 3 mm radius sapphire sphere against a sapphire plane, because both the normal load and the normal stiffness have values significantly lower than expected according to Hertz theory. For example, in the case of a contact between a sapphire plane and a sapphire sphere of radius 3mm, the expected theoretical value of the stiffness would be approximately $K_{z, th}=2.6 \cdot 10^6$ N/m for a penetration depth of 14 nm, to be compared with the much weaker experimental value $K_{z,\,exp}=.4.2\;10^4$ N/m.

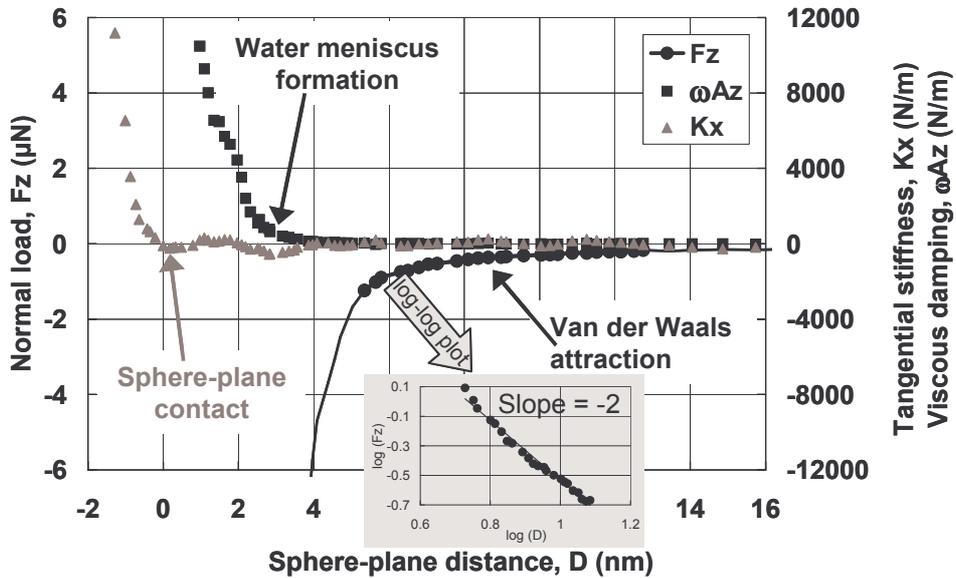

*Figure 3: Example of measured van der Waals attraction (black dots) for a type I experiment. The van der Waals attraction is measurable before the formation of the water meniscus, which is detectable from the viscous damping signal (black squares). The contact between the solids occurs at the beginning of the increase of the measured tangential stiffness (grey triangle).*

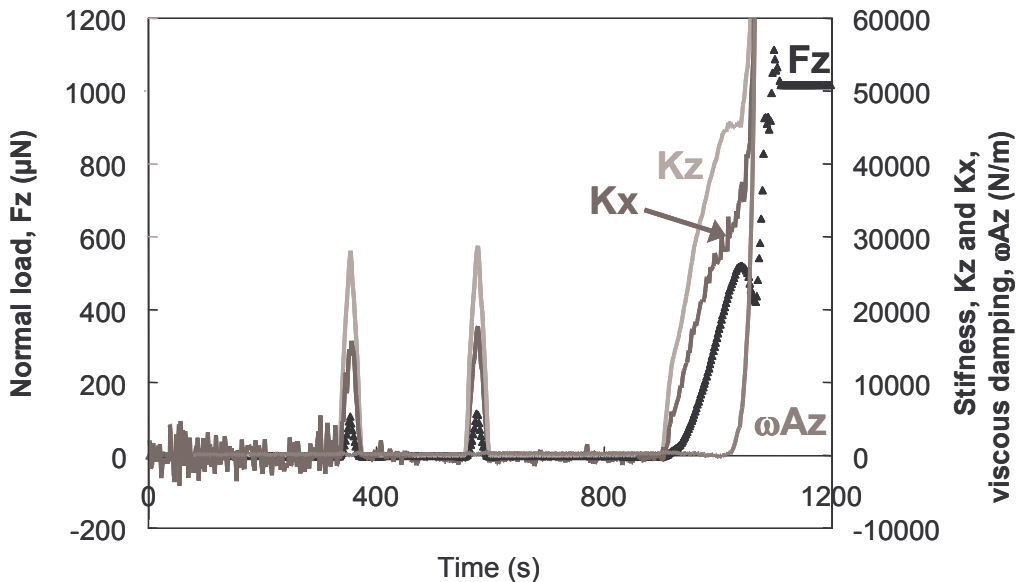

*Figure 4: Normal load $F_z$, normal stiffness $K_z$, tangential stiffness $K_x$, and viscous damping $\omega A_z$ versus time, for experiment of type II corresponding to the 3 first steps (ABC), (CDE), (EF) described on figure 1.*

As XPS analysis and nano-indentation tests (not shown in this paper) on the sapphire plane have demonstrated the absence of a soft metallic contamination surface layer, these low values did not come from lower surface mechanical properties. This has been then attributed to a contact area smaller than expected. Using Hertz calculations, it came that the measured normal load and stiffness was consistent with the contact between a sapphire plane and a sapphire sphere, with a radius in the micrometer range (see figure 5). Such a small radius was interpreted to be the radius of an asperity onto the sphere (this was observed on different points of the plane). This interpretation is consistent with the fact that, while the load remained lower than 500 µN, the viscous part of the dynamic signal was not significant, which shows that there was no macroscopic water meniscus in the contact.

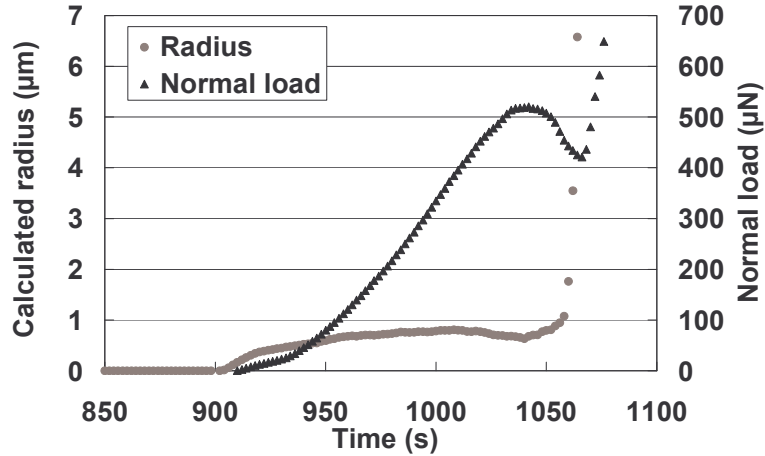

*Figure 5:* Calculated radius of the asperity (grey triangles) from the experimentally measured normal stiffness, using Hertz elastic theory, at the beginning of the loading stage (normal load inferior to 500 µN).

For larger loads (> 500 µN), before the nanofriction test, the normal load and stiffness increased very rapidly, so too did the apparent sphere radius. Simultaneously, some significant viscous behaviour was observed inside the contact, indicating that a macroscopic water meniscus has formed between the sphere and the plane. This resulted in an attractive peak at the same distance for both the loading and the unloading curves and a large hysteresis in the load/distance curve (figure 6).

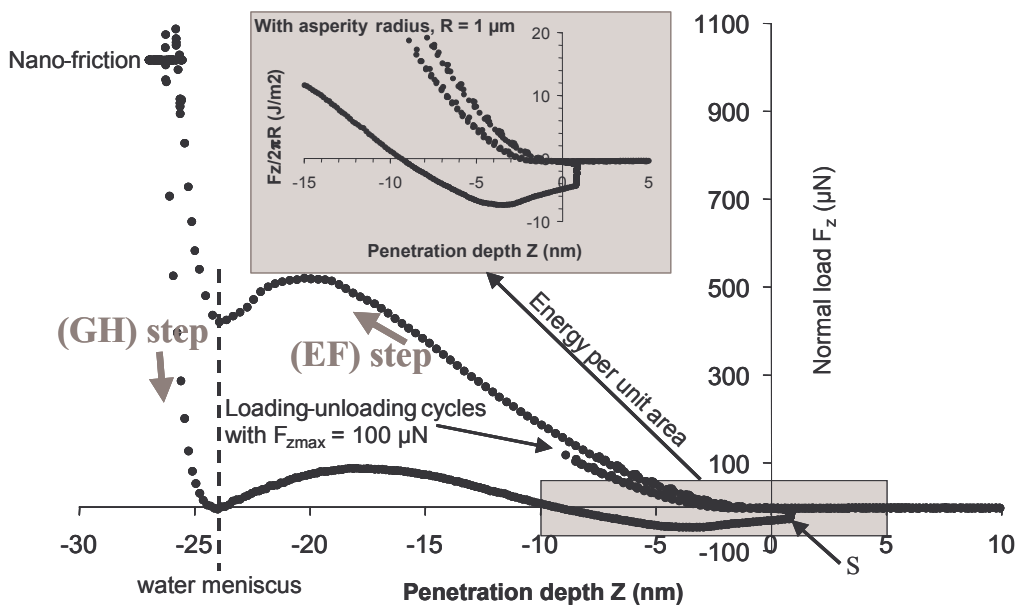

*Figure 6:* Normal load $F_z$ versus normal displacement curve for the test described on figure 1 -after nanofriction step (GS)-. In insert, calculation of the energy per unit area for the asperity-plane contact with $R_{asperity}=1µm$.

At the end of the unloading, (GH) step -as defined on figure 2-, before separation between the plane and asperity, another adhesion peak was observed which was attributed to electrostatic interaction between the surfaces. Indeed this peak could not be due to the meniscus attraction because there is no more damping. This interaction was not observed for similar experiments at the same maximum load (1000 μN) without the preliminary nano-friction step. For this part of the curve, the energy per unit area, using the Derjaguin, Muller, Toporov (DMT) model [17], $F_z/2\pi R$, was calculated taking R = 1 μm for the asperity radius (see insert figure 5) and was equal to -7.4 J/m$^2$ for the attraction peak. This very high value is in the same order of magnitude as the work of adhesion value calculated by Horn and Smith who measured the electrostatic force between dissimilar insulators after several contacts [18].

When the displacement was stopped and the sphere-plane distance maintained constant, the normal load $F_z$ decreased to zero after a few minutes.

### 3.2 Features of the charging

*3.2.1 Mobility of the charges*

As detailed in figure 7, at the end of the unloading, the electrostatic attraction decreased with time for an imposed sphere-plane distance (D=1nm). The experimental relaxation was well fitted by an exponential law $F_z = F_O \exp(-(t-t_0)/\tau)$, with $\tau = 256$ s. This shows that electrostatic charges generated by the nano-friction experiments are mobile on the sapphire surface.

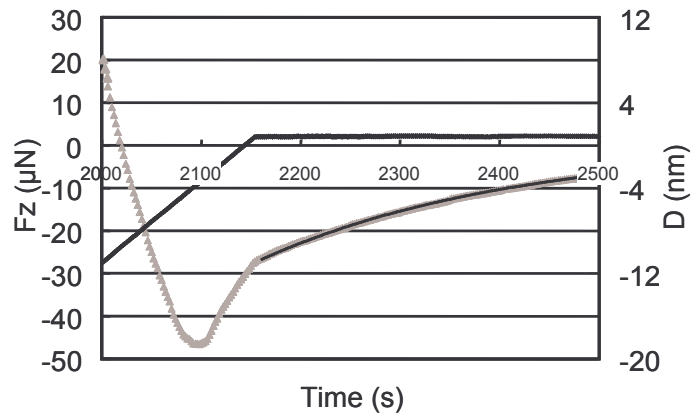

*Figure 7*: Normal load $F_z$ (grey triangles) and separation distance D (black curve) versus time during the unloading step (GH) for the test described on figure 1. The fine black curve corresponds to a decreasing exponential fit, with a time constant of 256 s.

The value of the relaxation time is in good agreement with the results of Felidj et al, who has found that the relaxation time of charges deposited on "non-freshly prepared" alumina surfaces is about 200 s [19]. They explained that the diffusion of the charges for these samples occurred exclusively on the surface due to a water layer present on the surface. In their case, the totality of the deposited charges disappeared when oxide surfaces have been exposed to ambient atmosphere. On the contrary, in our study, the charges generated by the nano-friction experiments did not disappear. Their presence was ascertained by the presence of a long-range attraction force in the nano-friction area during several days after the experiment. We have also observed these charges on scanning electron microscopy images, where they appear as white spot on the surfaces. We can suppose that the experimental cares to reduce the water presence, were not sufficient to remove all the water, but sufficient to obtain a partial coverage of the surface with water. In this case, the generated charges spread out only over a small area around the nano-friction area. This result is consistent with previous work on charge transfer between a metallic tip and metallic oxides during friction by St Jean et al, who observed very stable charges, but with a charge distribution larger than expected. These large dimensions were attributed to a possible diffusion on the surface assisted by the strong electric field between the tip and the surface [7].

Another explanation of this spreading of the electric charges is the mobility of charged dislocations. Indeed, in the case of alumina samples, the dislocations are known to serve as traps for electrons and to be mobile -when electrically charged- under an electric field [20]. Dislocations are present in our samples -due to the polishing and perhaps to the nano-friction test-. We assume that they could be electrically charged by the trapping of the charges generated by the nano-friction test, and that they could be mobile due to the electric field generated by these same charges. The mobility of charged dislocations could be in this case responsible of the charge mobility. As the charges spread, the electric field could decrease so that the mobility of the charged dislocations could diminish until they stop. In this case, the charges could be more stable than the one present in a water film, because they could be trapped either on the surface or in the bulk states. This second explanation is in good agreement with the mobility of the charges and their better stability evidenced by the experimental results.

In order to separate this relaxation effect from the force-displacement relationship during unloading, a quasi-instantaneous unloading procedure has been developed with fast recording of the resulting evolution of the normal load. The normal load was found to decrease when the sphere-plane plane distance decreased. To go further, the calculations of Burnham et al. have been used to describe this behavior and to quantify the corresponding charge densities. These calculations are given in the next parts.

*3.2.2 Model of Burnham, Colton and Pollock (BCP model) [21].*

This model was developed to evaluate the normal force between a charged tip -considered as a sphere- and a charged plane -considered as infinite- as a function of the charge densities $\sigma_t$ on the tip and $\sigma_p$ on the plane. Both charge densities are supposed to be uniform in the region of the tip.

The BCP calculation is based on two principles. The first one is to replace the effective charge densities $\sigma_t$ and $\sigma_p$ by their equivalent systems, as shown on figure 8. The equivalent system for the charged sphere is a single charge $Q_t$ located at distance A within the tip. Similarly, the electric field due to $\sigma_p$ can be represented by two charges $\pm Q_p$ positioned at $\pm Z$, where Z is very large. The second principle is to calculate the interaction between these equivalent charges by the image charge method because of the presence of several dielectric-air, air-dielectric interfaces [22].

For such a situation, Burnham et al give the magnitude of the interaction force. In our case, the expression can be simplified (similar materials with the same dielectric constant), and we thus obtain:

$$4\pi\varepsilon_0 F = -\frac{Q_t^2}{4(D+A)^2}\left(\frac{\varepsilon-1}{\varepsilon+1}\right) + \frac{RQ_tQ_p}{z(2D+A+R)^2}\left(\frac{\varepsilon-1}{\varepsilon+1}\right)^2$$

where D is the tip-plane distance, $\varepsilon$ the dielectric constant of sapphire ($\varepsilon \approx 10$), and R the radius of the tip.

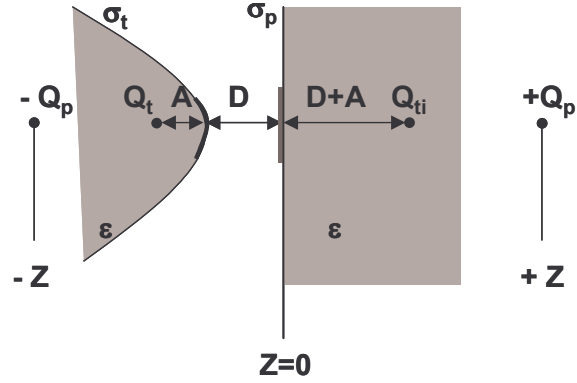

*Figure 8*: Schematic arrangement of a tip and a plane of same material (dielectric constant ε) in vacuum separated by a distance D, with their respective surface densities of charges $\sigma_t$ and $\sigma_p$. The equivalent electrostatic systems of these surface charges are respectively represented by a single charge $Q_t$ within the tip at distance A, and two opposite charges $\pm Q_p$ at large distance $\pm Z$ from the vacuum-plane interface. The image charge of the surface charge on the tip, due to dielectric-vacuum interface, is $Q_{ti}$ symmetric of $Q_t$ versus the interface.

The first term represents the interaction between the charge density on the tip surface, $\sigma_t$, with its image charge, i.e. with the sapphire polarisability. This interaction is always attractive. The second term describes the interaction between the charges present on the two antagonists. It can be attractive or repulsive depending on the respective signs of the charges. Note that for low tip-plane distance the first attractive term is always dominant (R>>D+A).

*3.2.3 Determination of the charge densities*

Using the BCP model, it was possible to fit the experimental force versus distance curve obtained after the nano-friction test, as shown on figure 9, providing that the considered radius R is the asperity one (R ≈ 1μm) and not the macroscopic one. In this case, the data were well fitted by the previous equation using the values: $Q_t \approx 5 \cdot 10^{-15}$ C, $Q_p/Z \approx 2 \cdot 10^{-7}$ C/m, A=26 nm. If we suppose that charges are generated only in the contact area, the value for $Q_t$ distributed over this contact area -estimated to be a disk of 2.24 μm radius from Hertz formula- yields to a charge density about $\sigma_t \approx 3 \cdot 10^{-4}$ C/m$^2$. Charges on the plane are of the same order. This simple model gives charge densities comparable with the values obtained by Burnham et al in the case of a diamond-graphite contact [21], and with the densities obtained by Sounilhac et al during the interaction of a tungsten tip with a $TiO_2$ plane [23]. The charge densities are an order of magnitude smaller than the one measured by Horn et al for mica-silica contact -about $3 \cdot 10^{-3}$ C/m$^2$-[10], and by Saint Jean et al after a friction experiment between a Pt coated tip and aluminium oxide -about $2 \cdot 10^{-3}$ C/m$^2$-[7]. A possible explanation for this smaller charge transfer is probably the similar work function of the two antagonists.

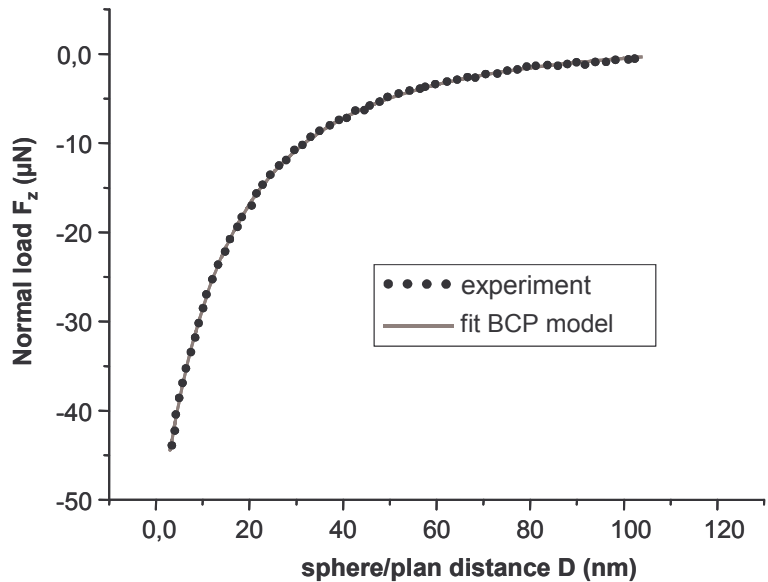

*Figure 9*: Normal load $F_z$ versus sphere-plane distance D during the quasi-instantaneous unloading step. Comparison between experimental data (black points) and the best fit with the equation deduced from the model of Burnham et al (grey line).

## 4. Conclusions

In this paper, the study of nano-friction experiments between two similar sapphire antagonists in sphere-plane contact under dry argon atmosphere was presented.
For very smooth surfaces:
    (i) the van der Waals interaction was first put in evidence and was characteristic of sapphire-sapphire contact in humid air.
    (ii) then at smaller sphere-plane distance, viscous damping appeared, corresponding to the formation of water meniscus, and is dominant, underlying the other contribution to the attraction.
For sphere with an asperity in the contact area (local higher contact pressure):
    (i) during the contact with the asperity, the viscous damping was negligible, and an additional attractive force was measured. This force was attributed to the generation of electrical charges (imaged by SEM) due to nano-friction, and was found to decrease with both time and distance.
    (ii) the electrical charges were found to be mobile on the surface. This mobility could be either attributed to a surface diffusion on water polluted alumina surfaces -identical relaxation time around 200 s-, or to the motion of electrically charged dislocations under the electric field due to the nano-friction generated charges. This initial mobility could explain the spread out of the charge over a localised area around the contact area.
    (iii) the distance dependence of this force was consistent, using the BCP model, with densities of charges of the order of $10^{-4}$ C/m$^2$.
In both cases, flat sphere and sphere with an asperity, the electrical charges were responsible for a long range attraction (measurable for values of D as large as 3µm) that could be still measured after several days.
    This study has evidenced the charge generation during the nano-friction of two samples of the same insulating material. However more work is necessary to identify the charge generation mechanism and its links with the density of available electronic states -i.e. sites of trapping- of the material.